\documentclass[12pt]{article}

\usepackage{amssymb, amsmath, amsthm}
\usepackage{graphicx}
\usepackage{cite}
\usepackage{hyperref}
\usepackage[left=2cm,right=2cm,top=2cm,bottom=2cm]{geometry}

\def\be{\begin{equation}}
\def\ee{\end{equation}}
\def\bea{\begin{eqnarray}}
\def\eea{\end{eqnarray}}

\newcommand{\td}{\text{d}}

\newtheorem{theorem}{Theorem}

\title{ \bf{All  higher-dimensional Majumdar-Papapetrou black holes}}

\author{James Lucietti\footnote{j.lucietti@ed.ac.uk } 
\\ \\ \small \sl School of Mathematics and Maxwell Institute for Mathematical Sciences, \\ \small \sl    University of Edinburgh, King's Buildings, Edinburgh, EH9 3FD, UK }

\date{}

\begin{document}

\maketitle

\begin{picture}(0,0)(0,0)
\put(350, 240){}
\put(350, 225){}
\end{picture}

\begin{abstract}  
We prove that the only asymptotically flat spacetimes with a suitably regular event horizon, in a generalised Majumdar-Papapetrou class of solutions to higher-dimensional Einstein-Maxwell theory, are the standard multi-black holes.  The proof involves a careful analysis of the near-horizon geometry and an extension of the positive mass theorem to Riemannian manifolds with conical singularities. This completes the classification of asymptotically flat, static, extreme black hole solutions in this theory.
\end{abstract}

\vspace{1cm}
The Majumdar-Papapetrou solution to Einstein-Maxwell theory represents the static equilibrium of an arbitrary number of charged black holes whose mutual electric repulsion exactly balances their gravitational attraction~\cite{Hartle:1972ya}. This remarkable configuration was later understood to arise as a supersymmetric solution to $\mathcal{N}=2$ supergravity, i.e., it saturates the BPS bound and admits Killing spinors~\cite{Gibbons:1982fy}. More recently it has been shown that it is in fact the only family of BPS black holes in this theory~\cite{Chrusciel:2005ve}.

In higher dimensions Einstein-Maxwell theory is not a consistent truncation of a supergravity theory. Nevertheless, asymptotically flat static solutions to the Einstein-Maxwell equations obey a BPS-like inequality $M\geq |Q| $ in all dimensions $n\geq 4$, where $M$ is the ADM mass and $Q$ is the Maxwell charge (in suitable units)~\cite{MuA, Kunduri:2017htl}. The $M>|Q|$ case has been fully solved by generalising the ingenious method of Bunting and Masood-ul-Alam~\cite{BMuA} to higher-dimensions, proving that the unique non-trivial regular solution is the non-extreme Reissner-Nordstr\"om black hole~\cite{Gibbons:2002ju, Kunduri:2017htl}.  

In this note we consider the extreme case $M=|Q|$, which implies the solution takes a `generalised' Majumdar-Papapetrou form~\cite{Kunduri:2017htl},
\be
g = - H^{-2} \td t^2 + H^{\frac{2}{n-3}} h_{AB} \td x^A \td x^B \; ,\qquad F = - \td H^{-1} \wedge \td t  \; ,  \label{MP}
\ee
where $\xi= \partial_t$ is the static Killing field and $(x^A)$ are coordinates on the orthogonal hypersurface $\Sigma$. Here $(\Sigma, h)$ is an $(n-1)$-dimensional Ricci-flat Riemannian manifold that is asymptotically-flat with zero ADM mass, and the function $H$ is harmonic on $(
\Sigma, h)$.  For  $n=4$ the  space $(\Sigma, h)$ is trivially flat, however in higher dimensions this need not be the case. We will perform a global analysis of this family of spacetimes for all dimensions $n\geq 4$ and determine the constraints imposed by the existence of a suitably regular event horizon.

The higher-dimensional  generalisation of the Majumdar-Papapetrou metrics was first found by Myers~\cite{Myers:1986rx} in the case $(\Sigma, h)$ is euclidean space. Heuristic arguments, analogous to those originally employed by Hartle and Hawking in four dimensions~\cite{Hartle:1972ya}, suggest that the only suitably regular solutions are the standard `multi-centre' solutions given by~\cite{Myers:1986rx},
\be
H= 1+\sum_{I=1}^N \frac{q_I}{r^{n-3}_I}   \label{multiBH}  \; ,
\ee
where $r_I = |x- p_I|$ is the euclidean distance from each centre $p_I \in \mathbb{R}^{n-1}$.  In four dimensions Hartle and Hawking demonstrated that these centres correspond to regular event horizons and these spacetimes can be analytically extended through these horizons~\cite{Hartle:1972ya}.  Curiously, in higher dimensions $n>4$, the solutions with multiple horizon components ($N>1$) do not have smooth horizons and analytic extensions do not exist in general~\cite{Welch:1995dh}. In particular, if $n=5$ the metric at the horizon is generically $C^2$ and the Maxwell field is $C^0$, whereas if $n>5$ the metric is generically $C^1$ at the horizon and the Maxwell field is $C^0$~\cite{Candlish:2007fh, Gowdigere:2014aca, Gowdigere:2014cqa, Kimura:2014uaa}. Therefore, as we explain below, we will allow for this lower differentiability in our analysis.

For dimension $n=4$  it has been proven that the only asymptotically flat regular black hole solutions in the Majumdar-Papapetrou class are the standard multi-black holes~\cite{Chrusciel:2006pc}, i.e., the harmonic function must take the multi-centre form (\ref{multiBH}). The proof requires detailed use of the near-horizon geometry.    In this note we show that a similar result holds in all dimensions: any suitably regular asymptotically flat black hole solution in the generalised Majumdar-Papapetrou class (\ref{MP}) must have (i) a base $(\Sigma, h)$ isometric to euclidean space (minus a point for each horizon) and (ii) a harmonic function $H$ of multi-centre type (\ref{multiBH}). Interestingly, the proof of (i) requires a mild extension of the positive mass theorem to  manifolds with conical singularities (we present this in the Appendix, as it may be of independent interest).  

More precisely, our main result is summarised in the following theorem:

\begin{theorem} 
Let $(\mathcal{M}, g, F)$ be a static, asymptotically flat, extreme solution to the  $n\geq 4$ dimensional Einstein-Maxwell equations such that:
\begin{enumerate}
\item The static Killing field $\xi$ is strictly timelike in the domain of outer communication $\langle \langle \mathcal{M} \rangle \rangle$ and null on the event horizon (and hence tangent to the null generators).
\item $(g,F)$ are smooth ($C^\infty$) in $\langle \langle \mathcal{M} \rangle \rangle$, whereas at the horizon: 
\begin{enumerate}
\item $g$ is $C^1$
\item $F$ is $C^0$ and the electric field $\iota_\xi F$ is $C^1$
\item  $(g,F)$ and derived-quantities are smooth in tangential directions.
\end{enumerate}
\item Each component of the horizon admits a smooth cross-section, i.e., an $(n-2)$-dimensional spacelike submanifold transverse to $\xi$, with an induced metric that is not Ricci-flat.
\end{enumerate}
Then, $(\langle \langle \mathcal{M} \rangle \rangle, g, F)$ is a Majumdar-Papapetrou solution (\ref{MP}) where the base $(\Sigma, h)$ is isometric to euclidean space with the points $p_{I=1, \dots, N} \in \mathbb{R}^{n-1}$ removed (each corresponding to a horizon component) and the harmonic function is of multi-centre form (\ref{multiBH}) with poles $p_I$.
\end{theorem}

Before presenting our proof it is helpful to make a few remarks to explain the rationale behind our assumptions. 
In fact,  assumption 1 has been proven for asymptotically flat, static, spacetimes with a  globally hyperbolic $\langle \langle \mathcal{M} \rangle \rangle$  under certain global assumptions~\cite{Chrusciel:2008rh}.
The regularity assumption 2 is required in order to capture the differentiability properties of the known multi-black hole solutions discussed above (assumption 2(c) is made for simplicity and could be relaxed).
Finally, assumption 3 together with our regularity assumption 2, allows one to introduce a precise notion of a near-horizon geometry~\cite{Kunduri:2013ana} that is also compatible with the black hole horizon topology theorems~\cite{Galloway:2011np}, which is crucial for deriving the geometry of $(\Sigma, h)$ near a horizon (assumption 2(b) concerning the electric field is required to control the subleading terms).  We will now present a proof of the above theorem. \\

\noindent {\it Proof.} As mentioned above, any asymptotically flat static solution to the Einstein-Maxwell equations that is extreme (in the sense $M=|Q|$) must be a Majumdar-Papapetrou solution (\ref{MP})~\cite{Kunduri:2017htl}. We first  record a number of spacetime invariants for this class which will be important in our analysis:
\bea
&&|\xi |^2 = -H^{-2}, \qquad \iota_\xi F= \td H^{-1} \; ,  \label{H} \\ 
&&\td \xi = 2 H^{-1} F \; . \label{dxi}
\eea
The minimal regularity for the static Killing field $\xi$ compatible with the  assumption that $g$ is $C^1$ at the horizon is that $\xi$ is also $C^1$ at the horizon. 
Then, from (\ref{H}) and assumptions 1 and 2, we deduce that the function $H^{-1}$ is positive and smooth in  $\langle \langle \mathcal{M} \rangle \rangle$,  and that $H^{-1}$ vanishes precisely at the event horizon and is $C^1$ at the horizon.  
It follows that 
\be
\td |\xi|^2 =- \td H^{-2}= -2 H^{-1} \td H^{-1}=0
\ee
on the event horizon, i.e., it is a degenerate Killing horizon of $\xi$.

Next we perform a careful near-horizon analysis. 
This is facilitated by assumption 3 which asserts that each component of the horizon admits a cross-section $S$. Then, the spacetime in a neighbourhood of a connected component of such a horizon can be written in Gaussian null coordinates $(x^\mu)=(v, \lambda, y^a)$ (see e.g.~\cite{Kunduri:2013ana}),
\be
g = - \lambda^2 f \td v^2+ 2 \td v \td \lambda + 2 \lambda h_a \td v \td y^a + \gamma_{ab} \td y^a \td y^b  \; ,  \label{gnc}
\ee
where $\xi = \partial_v$,   $(y^a)$ are coordinates on $S$, and $\lambda$ is an affine parameter for null geodesics transverse to the horizon synchronised so $\lambda=0$ on the horizon. Usually, the metric components are assumed to be smooth at and away from the horizon leading to the above form. Under our regularity assumptions the metric still takes the above form except now $f, h_a$ are $C^0$ and $\gamma_{ab}$ is $C^1$ at the horizon, with all components being smooth away from the horizon. We emphasise that $g_{vv}$ has a double zero at the horizon due to the fact that $H^{-1}$ is $C^1$ and vanishes at $\lambda=0$, together with $g_{vv}=-H^{-2}$.\footnote{ Thus, even though we assumed the metric is $C^1$ at the horizon, we deduce that $|\xi |^2$ is in fact $C^2$ at the horizon.}  

On the other hand, the minimal requirement for existence of the near-horizon limit --  defined by performing the diffeomorphism $(v, \lambda, y^a)\mapsto (v / \epsilon, \epsilon \lambda, y^a)$ and taking the limit $\epsilon \to 0$ -- is that $f, h_a, \gamma_{ab}$ are all $C^0$ at the horizon. Therefore, our assumptions still guarantee the existence of a near-horizon limit of the metric. The resulting near-horizon geometry takes the same form as (\ref{gnc}) with $f, h_a, \gamma_{ab}$ replaced by their values at $\lambda=0$, which in general we denote by $\mathring{f} \equiv f|_{\lambda=0}$ etc.  Note that our assumptions  imply that the near-horizon geometry itself is smooth, i.e., the data $\mathring{f}, \mathring{h}_a, \mathring{\gamma}_{ab}$ are a smooth function, 1-form and Riemannian metric on $S$. 

We now consider the Maxwell field. Normally, smoothness  (or at least $C^2$) of the solution is used to show that the near-horizon limit of the Maxwell field exists (in Gaussian null coordinates this requires $F_{va} = O(\lambda)$ and the rest of the components $O(1)$ near $\lambda=0$~\cite{Kunduri:2013ana}). However, given our lower regularity assumptions, it is not clear that a near-horizon limit of the Maxwell field exists in general.  For the Majumdar-Papapetrou class of solutions, the invariants (\ref{H}) imply
\be
H^{-1}= \lambda \sqrt{f} \; , \label{nhH}
\ee
and  $F_{v \mu} = (\iota_\xi F)_\mu=  \partial_\mu (\lambda \sqrt{f})$, where $f$ must be positive for small $\lambda>0$ (to ensure $\xi$ is timelike just outside the horizon).  Furthermore, using $\xi_\mu \td x^\mu = \td \lambda+ \lambda h_a \td y^a- \lambda^2 {f} \td v$, we find that (\ref{dxi}) gives
\be
F= - \td ( \sqrt{f}\lambda \td v )+ \frac{1}{2 \sqrt{f}\lambda} \td ( \lambda h_a \td y^a)  \; .   \label{Fgnc}
\ee
In particular, the $(\lambda a)$ component of (\ref{Fgnc}) gives $\partial_\lambda (\lambda h_a) = 2 \sqrt{f} F_{\lambda a} \lambda$, which  together with our assumption that $F,f, h_a$ are $C^0$ at the horizon, implies we can write $h_a= \lambda k_a$ where $k_a$ is $C^0$ at the horizon, i.e., $\mathring{h}_a=h_a|_{\lambda=0}=0$  and  $h_a$ is $C^1$ at the horizon.

It can be shown that staticity of (\ref{gnc}), i.e. that $\xi$ is hypersurface orthogonal, is equivalent to the following conditions~\cite{Figueras:2011va},
\bea
\partial_a f &=& \partial_\lambda (\lambda f) h_a - \lambda f \partial_\lambda h_a,   \label{stat1}\\
\partial_{[a} h_{b]} &=& h_{[a} \partial_\lambda(\lambda h_{b]}) \; . \label{stat2}
\eea
Using the above results, (\ref{stat1}) can be written as  $\partial_a f = \lambda v_a$ where $v_a :=  k_a \partial_\lambda (\lambda f)- f \partial_\lambda h_a$ is $C^0$ at the horizon. Evaluating this at $\lambda=0$ we immediately deduce that $f_0:=\mathring{f}$ is a constant on $S$, which must be non-negative (since $\xi$ is timelike for $\lambda>0$). Furthermore, it also follows that $\partial_a f$ is $C^1$ at the horizon.

We are now in a position to consider the near-horizon limit of the Maxwell field.  It is clear that the first term in (\ref{Fgnc}) has a well-defined near-horizon limit.  For the second term, we can write $F_{ab}= \partial_{[a} h_{b]}/ (2 \sqrt{f}) = \lambda^2 k_{[a} F_{|\lambda| b]}$, where in the second equality we have used the staticity condition (\ref{stat2}) and the explicit expression for $F_{\lambda a}$ given earlier. Therefore, since by assumption $F_{\lambda a}$ is $C^0$ at the horizon, it now follows that the near-horizon limit of the Maxwell field (\ref{Fgnc}) is simply\footnote{This argument is valid even if $\mathring{f}=0$, in which case $F_{\lambda a}|_{\lambda=0}= \lim_{\lambda \to 0} k_a /\sqrt{f}$ is finite.}
\be
F_{\text{\tiny NH}}=- \td ( \sqrt{f_0} \lambda \td v)  \; .  \label{FNH}
\ee
 Thus, despite our low regularity assumptions, the near-horizon limit of the Maxwell field still exists and we have a standard smooth near-horizon solution to the Einstein-Maxwell equations. Assumption 3 requires that the constant $f_0$ is positive, since if $f_0=0$ the near-horizon Maxwell field vanishes and the horizon metric $\mathring{\gamma}_{ab}$ is Ricci flat.
 
Even though we assumed that $F$ is $C^0$ at the horizon, the above analysis shows that the tangential components of the electric field $(\iota_\xi F)_a= \partial_a ( \lambda \sqrt{f})$ are $C^1$ at the horizon (in fact $C^2$). Therefore, since assumption 2(b) asserts the full electric field $\iota_\xi F$ is $C^1$, this reduces to requiring that the transverse component $(\iota_\xi F)_\lambda= \partial_\lambda( \lambda \sqrt{f})$ is $C^1$ at the horizon. 
  It then follows that $f$ is $C^1$ at the horizon.  In particular, this guarantees the existence of a well-defined first order correction to the near-horizon geometry as in~\cite{Li:2015wsa, Li:2018knr}, which will be helpful below.

The above near-horizon analysis  shows that, under our assumptions, the spacetime metric in a neighbourhood of a connected component of the horizon takes the form (\ref{gnc})
where  $f, h_a, \gamma_{ab}$ are $C^1$ at the horizon, $f_{\lambda=0}=f_0$ is a positive constant and $h_a =\lambda k_a$ for some $k_a$ which is $C^0$.  The orbit space metric is defined wherever $\xi$ is timelike and is given by $q_{\mu\nu} := g_{\mu\nu} - \xi_\mu \xi_{\nu}/ |\xi |^2$. Therefore, using (\ref{gnc}), we find that for $\lambda>0$,
\be
q = \frac{1}{f \lambda^2} ( d\lambda + \lambda^2 k_a dy^a)^2 + \gamma_{ab} dy^a dy^b  \; .  \label{nhorbit}
\ee
Observe that the horizon $\lambda=0$ is an infinite proper distance from any point, i.e., the orbit space $(\Sigma, q)$ is complete and a degenerate horizon corresponds to an asymptotic end even under our weak differentiability assumptions.\footnote{It is of course well known that for a  degenerate $C^2$ Killing horizon $(\Sigma,q)$ is a complete manifold such that any connected component of a degenerate horizon corresponds to an asymptotically cylindrical end (see e.g.~\cite{Chrusciel:1998rw, Khuri:2017fun}).}

On the other hand, the orbit space metric for (\ref{MP}) is simply $q= H^{\frac{2}{n-3}} h$. Comparing this to the general near-horizon orbit space metric (\ref{nhorbit}), we deduce that the base metric $h$ of (\ref{MP}) (which is invariantly defined where $\xi$ is timelike) near each component of the horizon can be written as
\be
h= \alpha^{-2}  f^{\alpha-1} [d\rho +  \alpha \rho^{n-2}  k_a dy^a]^2 + f^{\alpha} \rho^2 \gamma_{ab} dy^a dy^b  \; ,  \label{hnh}
\ee
where $\rho:= \lambda^{\frac{1}{n-3}}$ for $\lambda>0$,  $\alpha := 1/(n-3)$ and
\be
f = f_0+ O(\rho^{n-3}), \qquad \gamma_{ab} = \mathring{\gamma}_{ab}+ O(\rho^{n-3}), \qquad k_a= O(1)  \; ,  \label{nhfalloff}
\ee
as $\rho \to 0$. The order of the subleading terms is fixed by the fact that $f, h_a, \gamma_{ab}$ are $C^1$ at the horizon (as functions of $\lambda$).  In terms of the new coordinate
\be
H = \frac{1}{\sqrt{f} \rho^{n-3}}. \label{nhHrho}
\ee
To analyse the geometry of $(\Sigma, h)$ as $\rho\to 0$ it is convenient to adapt the near-horizon limit to this setting. Thus consider the diffeomorphism $\varphi_\epsilon: (\rho, y^a) \mapsto (\epsilon \rho, y^a)$ and define $h_\epsilon = \epsilon^{-2} \varphi^*_\epsilon h$. 
Then we find that as $\epsilon \to 0$  the 1-parameter family of metrics $h_\epsilon \to h_0$, where
\be
h_0 = \alpha^{-2} {f}_0^{\alpha-1} \left( d\rho^2 +\rho^2 \sigma_{ab}(y) dy^a dy^b  \right)  \;   \label{conical}
\ee
is a cone-metric of the compact space $(S, \sigma)$ defined by the horizon geometry
 \be
\mathring{\gamma}_{ab} \td y^a \td y^b  = \alpha^{-2} f^{-1} _0 \sigma_{ab}\td y^a \td y^b \; .  \label{horizonmetric}
\ee
Then, since $h_\epsilon$ is Ricci flat, it must be that $h_0$ is a Ricci flat cone-metric. It follows that $\sigma$ is an Einstein metric on $S$ normalised so $\text{Ric}(\sigma) = (n-3) \sigma$.  Defining $H_\epsilon= \epsilon^{n-3}\varphi^*_\epsilon H$ and using (\ref{nhHrho})  we find that $H_\epsilon\to H_0= 1/(\sqrt{f_0} \rho^{n-3})$ is automatically harmonic in the cone metric (\ref{conical}). Thus no further conditions on the near-horizon geometry occur for this class of solutions (\ref{MP}).\footnote{Alternatively,  (\ref{FNH}) and the near-horizon Einstein equation~\cite{Kunduri:2013ana} imply (\ref{horizonmetric}) where $\text{Ric}(\sigma)=(n-3)\sigma$.}

To summarise, we have found that the near-horizon geometry must be a direct product of  AdS$_2$ and an Einstein space $(S,\sigma)$ normalised as above,
\be
g_{\text{\tiny NH}} = - f_0 \lambda^2 \td v^2+ 2 \td v \td\lambda +  (n-3)^2 f_0^{-1} \sigma_{ab} \td y^a \td y^b \; ,
\ee
with Maxwell field (\ref{FNH}).
This in itself is a nontrivial result. In general, the classification of static near-horizon geometries in higher dimensions is an open problem and one can have non-trivial solutions which are warped products of AdS$_2$ and non-Einstein metrics $\mathring{\gamma}_{ab}$~\cite{Kunduri:2009ud}. Thus we have found that constraints arising from the Majumdar-Papapetrou solution rule out the possibility of non-trivial near-horizon geometries.\footnote{This was not properly taken into account in previous attempts at classifying static extreme black holes~\cite{Rogatko:2003kj, Rogatko:2006gg}.}   In particular, for $n=4$ the space $(S, \sigma)$ must be isometric to the unit round $S^2$, whereas for $n=5$ it must be locally isometric to the  unit round $S^3$.  However, for $n>5$ the horizon $(S, \sigma)$ need not be a space form, although  Myers's theorem shows that it must  be compact with a finite fundamental group. It is interesting to note that our near-horizon analysis did not assume compactness of $S$ as is often done, but instead this is an output of our analysis.

Importantly, equations (\ref{hnh}), (\ref{nhfalloff}) and (\ref{conical}) also show that any connected component of a horizon corresponds to a conically singular end of $(\Sigma, h)$. That is, there is an end  $E$ diffeomorphic to $(0, \rho_0) \times S$ with a metric which approaches a cone metric, i.e.,
\be
| h - h_0|_{h_0} = O(\rho^{\delta})  
\ee
as $\rho \to 0$ for some $\delta>0$, where $h_0$ is the cone metric  (\ref{conical}) of a compact Riemannian manifold $(S, \sigma)$, $| \cdot |_{h_0}$ is the norm defined by $h_0$ and $\rho \in (0, \rho_0)$. 
Specifically, our near-horizon analysis (\ref{nhfalloff}) shows that $\delta = n-3$, and also $|\mathring{\nabla}^s h|_{h_0}=O(\rho^{\delta-s})$ for $1\leq s \leq n-3$ where $\mathring{\nabla}$ is the metric connection of $h_0$. 

Now, we also know that $(\Sigma, h)$ is Ricci-flat and asymptotically-flat with zero mass.  For complete Riemannian manifolds the positive mass theorem would immediately imply $(\Sigma, h)$ must be isometric to euclidean space~\cite{Witten:1981mf, Bartnik}. However, the conically singular end implies that $(\Sigma, h)$ is not complete and therefore the standard positive mass theorem cannot be applied.  

Now suppose that $(\Sigma, h)$ is flat. 
Then,  it follows that $h_\epsilon$ and hence $h_0$ are also flat metrics. The latter condition is equivalent to $(S, \sigma)$ being a maximally symmetric space with positive curvature  $\text{Riem}(\sigma)_{abcd}= \sigma_{ac} \sigma_{bd}- \sigma_{ad} \sigma_{bc}$. Thus $(S, \sigma)$ is isometric to a quotient of the unit round sphere $S^{n-2}/\Gamma$ where $\Gamma$ is a discrete subgroup of $O(n-1)$. This implies that the end $E$ is diffeomorphic to $\mathbb{R}^{n-1}/\Gamma - \{ p \}$, where $p\in \mathbb{R}^{n-1}$ is a fixed point of $\Gamma$ that corresponds to the conically singular end. Thus, supposing we have $N$ conically singular ends corresponding to  $p_1, \dots, p_N \in \mathbb{R}^{n-1}$, we deduce that $(\hat{\Sigma}= \Sigma \cup \{ p_1, \dots, p_N \}, h)$,  is a flat orbifold.  By a generalisation of the Cartan-Hadamard theorem for orbifolds~\cite{orbifolds}, it follows that $(\hat{\Sigma}, h)$ must be isometric to a global quotient of euclidean space. However, as $(\hat{\Sigma}, h)$ is also asymptotically flat, this quotient must be trivial and hence $(\hat{\Sigma}, h)$ is isometric to euclidean space.   Thus, we deduce that $(\Sigma, h)$ must be isometric to euclidean space with $N$ points removed, that is, $\Sigma \cong \mathbb{R}^{n-1}-\{p_1, \dots, p_N\}$ and $h=\delta$ is the euclidean metric. It also follows that  $(S, \sigma)$  is isometric to the unit round sphere for each conically singular end.

Let $(x^i)$ be cartesian coordinates on $\mathbb{R}^{n-1}$ and $p \in \mathbb{R}^{n-1}$ correspond to a  horizon component. The coordinate change $(x^i) \mapsto (\rho, y^a)$ maps the euclidean metric to the  general form for the base metric  near the horizon (\ref{hnh}) if and only if
\bea
&&\partial_\rho x^i \partial_\rho x^i = \alpha^{-2} f^{\alpha-1}, \qquad \partial_\rho x^i \partial_a x^i = \alpha^{-1} f^{\alpha-1} \rho^{n-2} k_a , \nonumber \\ &&\partial_a x^i \partial_b x^i =\rho^2 f^\alpha \gamma_{ab}+ \rho^{2(n-2)} f^{\alpha-1} k_a k_b  \; .
\eea
In particular, this implies that $\partial_\rho x^i = O(1)$ and $\partial_a x^i = O(\rho)$  as $\rho \to 0$  and hence
\be
x^i- p^i = O(\rho) \;   . \label{cartnh}
\ee
We may now determine the precise singular structure of $H$ at a horizon.
Using (\ref{nhHrho}) and (\ref{cartnh}) we find that as $x\to p$, or equivalently as $\rho \to 0$,
\be
| x-p |^{n-3} H = \frac{1}{\sqrt{f}} \left( \frac{ | x-p |}{\rho} \right)^{n-3}= O(1)  \;. \label{Hsing}
\ee
Recall the harmonic function $H$ must be smooth in $\langle \langle \mathcal{M} \rangle \rangle$ and singular at the horizon. Therefore, in cartesian coordinates $H$ must have an isolated singularity at $x=p$. Hence (\ref{Hsing}) shows that $H$ has a pole of order $n-3$ at $x=p$. From the standard theory of harmonic functions in euclidean space we deduce that
\be
H = \frac{q_0}{|x-p|^{n-3}} + K
\ee
where $q_0$ is constant and $K$ is a harmonic function smooth in a neighbourhood of $x=p$.

We may now use this to derive global constraints on the spacetime via elementary arguments. 
Above we have shown that any connected component of a horizon corresponds to a pole of $H$  of order $n-3$ and $H$ is  smooth elsewhere. Thus, if the horizon has $N$-connected components corresponding to the points $x=p_I$, $I=1, \dots, N$, we can write
\be
H = \sum_{I=1}^N \frac{q_I}{r_I^{n-3}}  + \tilde{H}  \; ,
\ee
where $q_I$ are constants, $r_I= | x - p_I|$ and $\tilde{H}$ is a harmonic function which is smooth everywhere on $\mathbb{R}^{n-1}$.  Furthermore, asymptotic flatness requires $H \to 1$ as $r \to \infty$. Therefore $\tilde{H}$ is a bounded regular harmonic function on $\mathbb{R}^{n-1}$ and hence must be a constant. The constant is fixed by the asymptotics to be $\tilde{H}=1$ and hence we arrive at the general solution (\ref{multiBH}) corresponding to the standard multi-black hole solution. This completes the proof in the case $(\Sigma, h)$ is flat.  

To complete the proof, it remains to establish that $(\Sigma, h)$ must be flat. As discussed above, this  would follow from a generalisation of  the rigidity part of the positive mass theorem to conically singular manifolds.  In fact, for our purposes we only need the  rigidity part of the following generalisation of a simpler version of the positive mass theorem~\cite{Witten:1981mf, Bartnik}: {\it   Any asymptotically-flat Riemannian manifold $(\Sigma, h)$ with conical singularities and $\text{Ric}(h)\geq 0$, must have ADM mass $m\geq 0$ and $m=0$ if and only if  $(\Sigma, h)$ is flat.}\footnote{This theorem is only valid for `point-like' conical singularities as above. For higher-dimensional conical singularities it can  be false, e.g., the Eguchi-Hanson metric with angles identified so that it is asymptotically-euclidean gives a non-trivial zero-mass Ricci-flat metric with a conical singularity over a  bolt.}
We sketch a proof of this in the Appendix. Thus applying this to our case we deduce that $(\Sigma, h)$ is flat, which completes the proof.  $\Box$ \\ 

We close with a few remarks. The above analysis also classifies asymptotically-flat, static, supersymmetric black holes in five-dimensional minimal supergravity. This is because these must also take the form (\ref{MP}) with $(\Sigma, h)$ hyper-K\"ahler (and hence Ricci flat)~\cite{Gauntlett:2002nw}. In this case a different uniqueness proof has been previously given for supersymmetric (not necessarily static) black holes with a locally $S^3$ horizon, by assuming the supersymmetric Killing field is strictly timelike outside the black hole~\cite{Reall:2002bh}. In this context the conical singularity in the base is an ADE singularity which may be resolved to yield a complete asymptotically-flat hyper-K\"ahler base which therefore must be $\mathbb{R}^4$ (thus avoiding the need to invoke the positive mass theorem).  Our result also complements the recent classification of supersymmetric black holes with biaxial symmetry in five-dimensional minimal supergravity~\cite{Breunholder:2017ubu}. It would be interesting to complete the classification of supersymmetric black holes in this theory. 
 
This work may be viewed as an analogue of the   static black hole uniqueness proof of Bunting and Masood-ul-Alam~\cite{BMuA} for extreme black holes. Their method involves gluing two conformally rescaled copies of the orthogonal spatial hypersurface along the inner boundaries corresponding to the horizon, resulting in an asymptotically-flat zero-mass complete surface with non-negative scalar curvature, which by the positive mass theorem must be isometric to euclidean space. For extreme black holes we found this method does not work. Instead, a horizon manifests itself as a conical singularity of the asymptotically-flat  zero-mass Ricci-flat manifold $(\Sigma, h)$  (rather than a boundary) and  a mild generalisation of the positive mass theorem to accommodate such singularities is sufficient to establish it is flat.  It would be interesting to apply this theorem to prove similar  uniqueness results in other theories which support static extreme  black holes and  branes. \\
 
\noindent {\bf Acknowledgments.}   I would like to thank Marcus Khuri for helpful suggestions regarding a proof of the positive mass theorem required in this work. I am supported by a Leverhulme Trust Research Project Grant.

\appendix

\section{Positive mass  on manifolds with conical singularities}

Here we prove a generalisation of a simple version of the positive mass theorem due to Witten~\cite{Witten:1981mf} and Bartnik~\cite{Bartnik} to allow for conically singular ends, that was invoked in the main text. First, we recall the definitions of asymptotically flat and conically singular ends for a $d\geq 3$-dimensional Riemannian manifold $(\Sigma, h)$.

By definition~\cite{Bartnik}, an asymptotically flat end $E_\infty$ of $(\Sigma, h)$ is an end that is diffeomorphic to $\mathbb{R}^d \backslash B$ with $B$ a closed ball, where
\be
h_{ij} =   \delta_{ij} +O(r^{-\tau}) \; ,\qquad \partial_k h_{ij} =O(r^{-\tau-1})\; ,
\ee
as $r =\sqrt{x^i x^i}\to \infty$,  $(x^i)$ are cartesian coordinates on  $E_\infty$ defined by the diffeomorphism and $\tau>0$ is the decay rate.
 The ADM mass is
\be
m :=\lim_{r\to \infty}c_d  \int_{S_r} ( \partial_j g_{ji} - \partial_i g_{jj} ) \td S^i  \; ,
\ee
where $S_r$ is the sphere of constant $r$ in $E_\infty$, $c_d$ an irrelevant positive constant and the decay rate $\tau> (d-2)/2$ is required for $m$ to be well-defined~\cite{Bartnik}. 

On the other hand, we define a conically singular end $E_0$  of $(\Sigma, h)$ as follows: $E_0$ is diffeomorphic to $C= (0, \rho_0)\times S$ where $\rho_0>0$, $(S, \sigma)$ is a compact Riemannian manifold,
\be
|h-h_0|_{h_0}= O(\rho^\delta)  \; ,  \qquad   | \mathring{\nabla} h |_{h_0}= O(\rho^{\delta-1})  \; ,\label{hcone}
\ee
as $\rho \to 0$, the decay rate $\delta>0$, the norm $| \cdot |_{h_0}$ and connection $\mathring{\nabla}$ are with respect to the cone metric $h_0= \td \rho^2+ \rho^2 \sigma_{ab} \td y^a \td y^b$ on $C$, 
and the coordinates $(\rho, y^a)$ are defined by the diffeomorphism such that $\rho \in (0,\rho_0)$ (this is similar to other definitions of conical singularities~\cite{Lotay, Moore}).

We are now ready to state our result.

\begin{theorem}
 Let $(\Sigma, h)$ be a $d\geq 3$-dimensional asymptotically-flat Riemannian manifold with conical singularities. If $\text{Ric}(h)\geq 0$ then the ADM mass $m \geq 0$ and $m=0$ occurs iff $(\Sigma, h)$ is flat. 
 \end{theorem}

\noindent {\it Proof.} For simplicity of notation we will assume $(\Sigma, h)$ has one asymptotically-flat end $E_\infty$ and one conically singular end $E_0$, although the arguments below generalise to multiple ends straightforwardly. Thus, we assume there is a compact manifold $K$ such that $\Sigma - K = E_\infty \cup E_0$.

We follow closely the proof for the standard case where $(\Sigma, h)$ is a complete manifold~\cite{Witten:1981mf,Bartnik}. Thus suppose $z^i$ are globally defined harmonic functions on $(\Sigma,h)$ such that on the asymptotically flat and conically singular ends,
\bea
&&\partial^s (z^i-x^i)= O(r^{1-\tau-s}) \; , \qquad \text{as} \quad r\to \infty\;, \label{zinfty} \\ 
&&|\mathring{\nabla}^s(z^i- p^i)|_{h_0} = O(\rho^{\varsigma-s})\; , \qquad \text{as}\quad  \rho \to 0 \;,  \label{zcone}
\eea
respectively,  for $0\leq s \leq 2$ where the decay rate $\varsigma > 0$ is to be chosen at our convenience. In the absence of the conically singular end the proof that such harmonic functions exist was given by Bartnik~\cite{Bartnik}. In particular, these provide a set of cartesian coordinates at infinity known as harmonic coordinates. 

In the presence of a conically singular end, we may  construct such harmonic coordinates as follows. Let $y^i$ be harmonic coordinates on $E_\infty$ (guaranteed to exist by~\cite{Bartnik}) and extend these to $C^\infty(\Sigma)$ such that on $E_0$ they are constants $p^i$. Define $f^i:= -\Delta y^i$, where $\nabla$ and $\Delta= \nabla^A \nabla_A$ is the metric connection and Laplacian of $h$. Clearly $f^i$  vanishes identically on the ends and hence has compact support on $\Sigma$. Now consider the elliptic problem on $\Sigma$:
\be
\Delta v^i = f^i, \qquad \partial^s v^i  = O(r^{-\tau-s}), \qquad |\mathring{\nabla}^s v^i|_{h_0} = O(\rho^{\varsigma-s}) \; ,
\ee
where $f^i$ is fixed as above.
By the maximum principle,  a solution $v^i$ to this system is unique since $v^i\to 0$ in both ends. Then, defining $z^i := y^i+v^i$ gives a set of harmonic functions on $\Sigma$ which obey the decay rates (\ref{zinfty}) and (\ref{zcone})  (the former follows from $\partial^s(y^i-x^i)= O(r^{1-\tau-s})$~\cite{Bartnik}). To establish existence of a Green's function for this problem rigorously one could presumably adapt the arguments in~\cite{Bartnik}, perhaps using the theory for  the Laplacian on manifolds with admissible metrics  (which include both asymptotically-flat and conically singular ends)~\cite{Lockhart}.   We will not pursue this here.

Now, define the 1-forms $K^i= \td z^i$ which, in view of the $z^i$ being harmonic, must obey the Bochner identity 
\be
\Delta | K^i |^2 =2 | \nabla K^i |^2 + 2\, \text{Ric}(K^i, K^i)
\ee
for each $i=1, \dots, d$. Integrate this over $\Sigma$ to deduce
\be
\lim_{r \to \infty} \int_{S_r} \partial_j  |K^i |^2 \td S^j - \lim_{\rho\to 0} \int_{S_\rho} \partial_n  |K^i |^2 \td \text{vol}  = 2\int_{\Sigma} \big[  | \nabla K^i |^2 + \text{Ric}(K^i, K^i) \big] \td \text{vol} \geq 0  \; ,  \label{intBoch}
\ee
where $S_\rho$ is a surface of constant $\rho$ in $E_0$ and $n$ is the unit-normal to $S_\rho$.  An important property of  harmonic coordinates is that in terms of them the ADM mass simplifies to
\be
 m =-\frac{c_d}{2} \int_{S_\infty} \partial_j g_{ii}  \td S^j =  \sum_{i=1}^d \frac{c_d}{2} \int_{S_\infty} \partial_j |K^i |^2  \td S^j   \; ,
\ee
where to obtain the second equality we have used the fact that in the  harmonic coordinates $(z^i)$ we have $| K^i |^2 = g^{ii}$ (no sum).
On the other hand, for the integral near the conical singularity we find that (\ref{zcone}) and (\ref{hcone}) imply,
\be
I_\rho:= \int_{S_\rho} \partial_n  |K^i |^2 \td \text{vol}  = O(\rho^{2\varsigma +d-4}) \; .   \label{coneint}
 \ee
 To see this  first note that $| I_\rho| \leq | \partial_n G |_{\text{max}} \text{vol}(S_\rho)$ where we have set $G:=| K^i|^2$. Now, writing $G=G_0+(G-G_0)$ where $G_0:=|K^i|_{h_0}^2$ we have the bound $|\partial_n G|\leq |\mathring{\nabla} G_0|_{h_0}+|\mathring{\nabla} (G-G_0)|_{h_0}$. 
The first term is bounded by $|\mathring{\nabla} G_0|_{h_0}\leq 2 | \mathring{\nabla}\mathring{\nabla} z^i|_{h_0} |\mathring{\nabla} z^i|_{h_0} = O(\rho^{2\varsigma-3})$   using (\ref{zcone}), whereas the second term 
\be
|\mathring{\nabla} (G-G_0)|_{h_0} \leq 2 | h-h_0|_{h_0} | \mathring{\nabla}\mathring{\nabla} z^i|_{h_0} |\mathring{\nabla} z^i|_{h_0} + | \mathring{\nabla}(h-h_0)|_{h_0} |\mathring{\nabla} z^i|_{h_0} ^2 = O (\rho^{2\varsigma-3 +\delta})
\ee using (\ref{zcone}) and (\ref{hcone}). Thus, since $\delta>0$, we deduce that $|\partial_n G | = O(\rho^{2 \varsigma-3})$, which together with the fact that $\text{vol}(S_\rho) = O(\rho^{d-1})$, gives the result (\ref{coneint}).  

Therefore, (\ref{coneint}) vanishes as $\rho\to 0$ provided $2\varsigma+d -4 > 0$. For $d\geq 4$ this is trivially satisfied since $\varsigma>0$, whereas for $d=3$ this can be ensured by taking $\varsigma>(d-2)/2$. Thus, summing (\ref{intBoch}) over  $i=1, \dots, d$, we deduce that $m\geq 0$ with equality iff $\nabla K^i=0$ for all  $i=1, \dots, d$.   If $m=0$, the $K^i$ are parallel $1$-forms that form an orthonormal basis at infinity, which implies the $K^i$ are a global parallel orthonormal frame and hence $(\Sigma, h)$ is flat.  $\Box$

\end{document}